\documentstyle[12pt,fleqn]{article}

\begin{document}

\title{Developments in General Relativity: Black Hole Singularity 
and Beyond}
\author{I.D.Novikov\\
Theoretical Astrophysics Center\\ 
Juliane Maries Vej 30, DK-2100 Copenhagen,  Denmark;\\
University Observatory\\
Juliane Maries Vej 30, DK-2100 Copenhagen,  Denmark;\\
Astro-Space Center of Lebedev Physical Institute\\
Profsoyuznaya 84/32, Moscow 117997, Russia;\\
NORDITA, Blegdamsvej 17,  DK-2100 Copenhagen,  Denmark}

\maketitle

\begin{abstract}
At the 20-th Texas Symposium on Relativistic Astrophysics there was a 
plenary talk devoted to the recent developments in classical Relativity. 
In that talk the problems of gravitational collapse, collisions of black 
holes, and of black holes as celestial bodies were discussed. But probably 
the problems of the internal structure of black holes are a real great 
challenge. In my talk I want to outline the recent achievements in our 
understanding of the nature of the singularity (and beyond!) inside a 
realistic rotating black hole. This presentation also addresses the 
following questions:\\
Can we see what happens inside a black hole?\\
Can a falling observer cross the singularity without being crushed?\\
An answer to these questions is probably ``yes''.
\end{abstract}

\section{Introduction}
The problems of the internal structure of black holes are a real great
challenge. Inside a black hole the main sights are the singularity.

In this paper I will give a retrospective review and outline the recent 
achievements in our understanding of the nature of the singularity (and 
beyond) inside a realistic, rotating black hole.

I want to note that there is a strong believe that the interior of black 
holes is not visible for an external observer, and who cares for 
invisible things?

Natural question arises: Can one see from outside what happens inside a 
black hole? My presentation also addressed this question. An answer to 
this question is probably affirmative.

For systematic discussion of the problems of the internal structure of 
black holes see [1-8]. 

The problem of black holes interior was the subject of a very active 
investigation last decades. There is a great progress in these researches. 
We know some important properties of the realistic black hole's interior, 
but some details and crucial problems are still the subject of much 
debate.

A very important point for understanding the problem of black hole's 
interior is the fact that the path into the gravitational abyss of the 
interior of a black hole is a progression in time. We recall that inside 
a spherical black hole, for example, the radial coordinate is timelike. 
It means that the problem of the black hole interior is an {\it 
evolutionary problem}. In this sense it is completely different from a 
problem of an internal structure of other celestial bodies, stars for 
example, or planets.

In principle, if we know the conditions on the border of a black hole 
(on the event horizon), we can integrate the Einstein equations in time 
and learn the structure of the progressively deeper layers inside the 
black hole. Conceptually it looks simple, but there are two types of
principal difficulties which prevent realizing this idea consistently.

The first difficulty is the following. Formally the internal structure of 
a generic rotating black hole even soon after its formation depends 
crucially on the conditions on the event horizon at very distant future 
of the external observer (formally at the infinite future). This happens 
because the light-like signal can propagate from the very distant future 
to those regions inside a black hole which are deep enough in the hole. 
The limiting light-like signals which propagate from (formally) infinite 
future of the external observer form a border inside a black hole which 
is called a { \it Cauchy horizon}.

Thus, the structure of the regions inside a black hole depends crucially 
on the fate of the black hole at infinite future of an external observer. 
For example, it depends on the final state of the black hole quantum 
evaporation (because of the Hawking radiation), on possible collisions of 
the black hole with other black holes, or another bodies, and it depends 
on the fate of the Universe itself. It is clear that theoreticians feel 
themselves uncomfortable under such circumstances.

However we will see that all these uncertainties are related with the 
structure of spacetime of a realistic black hole very close to the 
singularity inside it.These parts of spacetime are in the region with the 
curvature greater than the Planck value. Probably the Classical General 
Relativity is not applicable here (see below, section 6).

The second serious problem is related to the existence of a singularity 
inside a black hole. A number of rigorous theorems (see references in [2])
imply that singularities in the structure of spacetime develop inside 
black holes. Unfortunately these theorems tell us practically nothing 
about the locations and the nature of the singularities. It is widely 
believed today that in the singularity inside a realistic black hole the 
characteristics of the curvature of the spacetime tends to infinity. 
Close to the singularity, where the curvature of the spacetime approaches 
the Plank value, the Classical General Relativity is not applicable. We 
have no a final version of the quantum theory of gravity yet, thus any 
extension of the discussion of physics in this region would be highly 
speculative. Fortunately, as we shall see, these singular regions are deep 
enough in the black hole interior and they are { \it in the future } with 
respect to overlying and { \it preceding} layers of the black hole where 
curvatures are not so high and which can be described by well-established 
theory.

The first attempts to investigate the interior of a Schwarzschild black 
hole have been made in the late 70's [9,10]. It has been demonstrated that 
in the absence of external perturbations at late times, those regions of 
the black hole interior which are located long after the black hole 
formation are virtually free of perturbations, and therefore it can be 
described by the Schwarzschild geometry for the region with radius less 
than the gravitational radius. This happens because the gravitational
radiation from aspherical initial excitations becomes infinitely diluted 
as it reaches these regions. But this result is not valid in general case 
when the angular momentum or the electric charge does not vanish. The 
reason for that is related to the fact that the topology of the interior 
of a rotating or/and charged black hole differs drastically from the 
Schwarzschild one. The key point is that the interior of this black hole
possesses a { \it Cauchy horizon }. This is a surface of infinite 
blueshift. Infalling gravitational radiation propagates inside the black 
hole along paths approaching the generators of the Cauchy horizon, and the 
energy density of this radiation will suffer an infinite blueshift as it 
approaches the Cauchy horizon.

This infinitely blueshifted radiation together with the radiation scattered 
on the curvature of spacetime inside a black hole leads to the formation 
of curvature singularity instead of the regular Cauchy horizon.

We will call this singularity Cauchy horizon singularity. A lot of papers 
were devoted to investigation of the nature of this singularity. In 
addition to the papers mentioned above see also [11-26]. Below we consider
main processes which are responsible for the formation of the singularity.

\section{Processes which are responsible for the formation of the Cauchy 
horizon singularity}

This section discusses the nonlinear effects which trigger the formation 
of a singularity at the Cauchy horizon inside a black hole. In the 
Introduction we emphasized that the problem of the black hole interior is 
an evolutionary problem, and it depends on the initial conditions at the 
surface of the black hole for all momenta of time up to infinity. To 
specify the problem, we will consider an isolated black hole (in 
asymptotically flat spacetime) which was created as a result of a 
realistic collapse of a star without assumptions about special symmetries.

The initial data at the event horizon of an isolated black hole, which 
determine the internal evolution at fairly late periods of time, are known 
with precision because of the no hair property.  Near the event horizon 
we have a Kerr-Newman geometry perturbed by a dying tail of gravitational
 waves. The fallout from this tail produces an inward energy flux 
decaying as an inverse power $v^{-2p}$ of advanced time $v$, where 
$p=2l+3$ for multipole of order $l$ see [27-30]. See details in Section 3.

Now we should integrate the Einstein equations with the known boundary 
conditions  to obtain the internal structure of the black hole. In 
general, the evolution with time into the black hole depths looks as 
follows. The gravitational radiation penetrating the black hole and partly
backscattered by the spacetime curvature can be considered, roughly 
speaking, as two intersecting radial streams of infalling and outgoing 
gravitational radiation fluxes, the nonlinear interaction of which leads 
to the formation a non-trivial structure of the black hole interior.
However in such a formulation it is a very difficult and still not solved 
completely mathematical problem. We will consider main achievements  in 
solving it.

What are the processes responsible for formation of the Cauchy horizon 
singularity? The key factor producing its formation is the infinite 
concentration of energy density close to the Cauchy horizon as seen by a
free falling observer. This infinite energy density is produced by the
ingoing radiative ``tail''.

The second important factor here is a tremendous growth of the black hole 
internal mass parameter, which was dubbed {\it mass inflation} [31].

We start by explaining the mechanism responsible for the mass inflation 
[32-34]. Consider a concentric pair of thin spherical shells in an empty 
spacetime without a black hole [35]. One shell of mass $m_{con}$ 
contracts, while the other one of mass $m_{exp}$ expands. We assume that 
both shells  are moving with the speed of light (for example, ``are made 
of photons''). The contracting shell, which initially has a radius 
greater than the expanding one, does not create any gravitational effects 
inside it, so that the expanding shell does not feel the existence of the 
external shell. On the other hand, the contacting shell moves in the 
gravitational field on the expanding one. The mutual potential of the 
gravitational energy of the shells acts as a debit (binding energy) on 
the gravitational mass energy of the external contracting shell.  Before 
the crossing of the shells, the total mass of both of them, measured by 
an observer outside both shells, is equal to $m_{con}+m_{exp}$ and is 
constant because the debit of the numerical increase of the negative 
potential energy is exactly balanced by the increase of the positive 
energies of photons blueshifted in the gravitational field of the 
internal sphere.

When shells cross one another, at radius $r_0$, the debit is transferred 
from the contracting shell to the expanding one, but the blueshift of 
the photons in the contraction shell survives. As a result, the masses of 
both spheres change. The increase of mass $m_{con}$ is called mass 
inflation. The exact calculation shows that the new mass $m'_{con}$ and 
$m'_{exp}$ are
\begin{equation}
m'_{con}=m_{con}+\frac{2m_{con}m_{exp}}{\varepsilon},\hspace{2cm}
m'_{exp}=m_{exp}-\frac{2m_{con}m_{exp}}{\varepsilon}
\end{equation}
where $\varepsilon\equiv(r_0-2m_{exp})$, (in the paper we use the units 
$G=1$, $c=1$). The total mass-energy is, of course, conserved 
$m'_{con}+m'_{exp}=m_{con}+m_{exp}$. If $\varepsilon$ is small (the
encounter is just outside the horizon of $m_{exp}$), the inflation of 
mass of $m_{con}$ can become arbitrary large. 

It is not difficult to extend this result to the shells crossing inside
a black hole. For simplicity consider at the beginning a spherical charged
black hole.

Ori [18] considered a continuous influx (imitating the ``tail'' of ingoing 
gravitational radiation) and the outflux as a thin shell (a very rough 
imitation of the outgoing gravitational radiation scattered by the 
spacetime curvature inside a black hole). He specified the mass 
$m_{in}(v)$ to imitate the Price power-law tail (see section 3) and found 
that the mass function diverges exponentially near the Cauchy horizon as 
a result of the ingoing flux with the outgoing crossing of the ingoing 
shell:
\begin{equation}
m\sim e^{k_0v_-}(k_0v_-)^{-2p}, \hspace{2cm} v_-\rightarrow\infty,
\label{q2}
\end{equation}
where $v_-$ is the advanced time in the region lying to the past of 
the shell, $k_0$ is constant, the positive constant $p$ depends on 
perturbations under discussions. Expression (2) describes {\it mass 
inflation}. In this model, we have a scalar curvature singularity since 
the Weyl curvature invariant $\Psi_2$. (Coulomb component) diverges 
at the Cauchy horizon. Ori [18] emphasizes that in spite of this 
singularity, there are coordinates in which the metric is finite at the 
Cauchy horizon.  He also demonstrated that though the tidal force in the 
reference frame of a freely falling observer grows infinitely its 
action on the free falling observer is rather modest. According to [18] 
the rate of growth of the curvature is proportional to  
\begin{equation}
\sim \tau^{-2}\left|\ln|\tau|\right|^{-2},
\label{q3}
\end{equation}
where $\tau$ is the observer's proper time, $\tau=0$ corresponds to the 
singularity. Tidal forces are proportional to the second time derivatives 
of the distances between various points of the object. By integrating the 
corresponding expression twice, one finds that as the singularity is 
approached $(\tau=0)$, the distortion remains finite.

There is one more effect caused by the outgoing flux. This is the 
contraction of the Cauchy horizon (which is singular now) with retarded 
time due to the focusing effect of the outgoing shell-like flux. This 
contraction continuous until the Cauchy horizon shrinks to $r=0$, and a 
stronger singularity occurs. Ori [18] has estimated the rate of approach 
to this strong singularity $r=0$.

In the case of realistic rotating black hole both processes the infinite 
concentration of the energy density and mass inflation near the Cauchy 
horizon also play the key role for formation of the singularity. 

\section{Decay of physical fields along the event horizon of isolated 
black holes}

Behavior of physical fields along the event horizon determines dynamics of 
these fields inside a black hole and has an impact on the nature of the 
singularity inside the black hole. We will consider here the behavior of
perturbations in the gravitational field. The first guess about the decay
of gravitational perturbations outside Schwarzschild black holes was given
in [36], a detailed description was given by Price [27,37]. 
References for subsequent work see in [2] and in the important work [38].
According to [27], any radiative multipole mode $l,m$  of any initially
compact linear perturbation dies off outside a black hole at late time
as $t^{-2l-3}$. The mechanism which is responsible for this behavior is
the scattering of the field off the curvature of spacetime asymptotically 
far from the black hole.

In the case of a rotating black hole the problem is more complicated due
to the lack of spherical symmetry. This problem was investigated in many 
works. See analytical analyses, references and criticism in [38], 
numerical approach in [39].
For individual harmonics there is a power-law decay which is similar to 
the Schwarzschild case except that at the event horizon the perturbation 
also
oscillates in the Eddington coordinate $v$ along the horizon's null
generators proportional to
\begin{equation}
\sim e^{im\Omega_+v},
\label{q4}
\end{equation}
where $\Omega_+$ is the angular velocity of the black hole rotation, 
$\Omega_+=a/2Mr_+$, $M$ and $a$ are mass and specific angular momentum 
of a black hole correspondingly, $r_+=M+\sqrt{M^2-a^2}$. Another important 
difference from the Schwarzschild case is the following. In the case of the
rotating black holes spherical-harmonic modes do not evolve 
independently. In the linearized theory there is a coupling between 
spherical harmonics multipole of different $l$, but with the same $m$. 
In the case fully nonlinear perturbations there is the guess that $m$ will 
not be conserved also. So in the case of arbitrary perturbation the modes
 with all $l$ which are consistent with the spin weight $s$ of the field 
will be exited. For the field with spin weight $s$ all modes with 
$l\ge|s|$ will be exited. Accordingly, the late-time dynamics will be 
dominated by the mode with $l=|s|$. The falloff rate is then 
$t^{-(2|s|+3)}$. In the case of the gravitational field it corresponds to 
$|s|=2$ and $t^{-7}$.

\section{Nature of the singularity of a rotating black hole}

As we mentioned in Section 2, we can use the initial data at the event
horizon which we discussed in the previous Section 3 to determine the
nature of the singularity inside a black hole. Main processes which are
responsible for formation of the singularity were discussed in Section2.

We start from the discussion of the singularity which arises at late time, 
long after the formation of an isolated black hole.

In general  the evolution with time into the black hole deeps looks like 
the following. There is a weak flux of gravitational radiation into a 
black hole through the horizon because of small perturbations outside of 
it. When this radiation approaches the Cauchy horizon it suffers an 
infinite blueshift. The infinitely blueshift radiation together with the 
radiation scattered by the curvature of spacetime inside the black hole 
results in a tremendous growth of the black hole internal mass parameter 
(``mass inflation'', see Section 2) and finally leads to formation of the 
curvature singularity of the spacetime along the Cauchy horizon. The 
infinite tidal gravitational forces arise here. This result was confirmed 
by considering different models of the ingoing and outgoing fluxes in the 
interior of charged and rotating black holes([6],[4]). 

In the case of a rotating black hole the growth of the curvature (and mass
function) when we coming to the singularity is modulated by the infinite
number of oscillations. This oscillatory behavior of the singularity is
related to the dragging of the inertial frame due to rotation of a black 
hole. Ori [4] calculated the asymptotic of the curvature scalar 
$K\equiv R_{iklm}R^{iklm}$ near the  Cauchy horizon singularity:
\begin{equation}
K\approx A|\tau|^{-2}\left[\ln (-\tau/M)\right]^{-7}
\times \sum_{m=1,2}C_m\exp\left\{-im\left[a(M^2-a^2)^{-1/2}\ln (-\tau /M)
\right]\right\},
\label{q5}
\end{equation}
where $\tau$ is a proper time of a free falling observer, $\tau=0$ at 
the singularity, $A$ is a constant that depends on the geodesic's 
constants of motion, $C_m$ are coefficients that depend on point at the 
singularity which is hit by the observer.

It was shown [18] that the singularity at the Cauchy horizon is quite 
weak. In particular, the integral of the tidal force in the freely falling 
reference frame over the proper time remains finite. It means that the 
infalling object would then experience the finite tidal deformations which
(for typical parameters) are even negligible. While an infinite force is 
extended, it acts only for a very short time. This singularity exists in a 
black hole at late times from the point of view of an external observer, 
but the singularity which arises just after the gravitational collapse of 
a star is much stronger. 
According to the Tipler's terminology [41] (see also generalization of the 
classification in [42]) this is a weak singularity.
It seems likely that an observer falling into a 
black hole with the collapsing star encounters a crushing singularity
(strong singularity in the Tipler's classification). This is so called 
Belinsky-Khalatnikov-Lifshitz (BKL) space-like singularity [40]. 
On the other hand  an observer falling into an isolated black hole in a 
late times generally reaches a weak singularity described above.

The weak Cauchy horizon singularity arises first at very late time 
(formally infinite time) of the external observer and its null generators
propagate deeper into black hole and closer to the event of the 
gravitational collapse. They are subject of the focusing effect under the 
action of the gravity of the outgoing scattered radiation (see Section 2).
Eventually the weak null singularity shrinks to $r=0$, and strong BKL 
singularity occurs. This picture was considered in details in the case of 
a charged spherical black hole but I do not know the strict proof of it 
in the case of a rotating black hole [4,25]

There is one more important question: how generic are considering 
singularities? The solution which describes a singularity is called 
general if it depends on the total number of arbitrary functions of three
independent variables which corresponds to the inherent degrees of freedom 
(plus the number of unfixed gauge degrees of freedom). In the case of 
gravitational field in vacuum this inherent degrees of freedom are equal 
4.

The authors of [40] have had demonstrated that BKL singularity is general.
Ori and Flanagan [21] have demonstrated that weak Cauchy horizon 
singularity is general also. So on principle both of these types of 
singularity are stable and can arise inside black holes. But of course 
which singularity (or both) arises depends on concrete situation, and 
this problem should be analyzed.

\section{Quantum effects}

As we mentioned in Introduction quantum effects play crucial role in the 
very vicinity of the singularity. In addition to that the quantum 
processes probably are important also for the whole structure of a black 
hole. Indeed, in the previous discussion we emphasized that the internal 
structure of black holes is a problem of evolution in time starting from 
boundary conditions on the event horizon for all moments of time up to 
the infinite future of the external observer.

It is very important to know the boundary conditions up to infinity 
because we observed that the essential events - mass inflation and 
singularity formation - happened along the Cauchy horizon which brought 
information from the infinite future of the external spacetime. However, 
even an isolated black hole in an asymptotically flat spacetime cannot 
exist forever.  It will evaporate by emitting Hawking quantum radiation. 
So far we discussed the problem without taking into account this ultimate 
fate of black holes. Even without going into details it is clear that 
quantum evaporation of the black holes is crucial for the whole problem.

What can we say about the general picture of the black hole's interior 
accounting for quantum evaporation? To account for the latter process we 
have to change the boundary conditions on the event horizon as compared to 
the boundary conditions discussed above. Now they should include the flux 
of negative energy across the horizon, which is related to the quantum 
evaporation. The last stage of quantum evaporation, when the mass of the 
black hole becomes comparable to the Plank mass 
$m_{Pl}=(\hbar c/G)^{1/2}\approx 2.2\times 10^{-5} g$, is unknown. 
At this stage the spacetime curvature near the horizon reaches 
$l_{Pl}^{-2}$, where $l_{Pl}$ is the Plank length:
\begin{equation}
l_{Pl}=\left(\frac{G\hbar}{c^{3}}\right)^{1/2}\approx 
1.6\times 10^{-33} {\mbox {cm}}.
\label{q6}
\end{equation}
This means that from the point of view of semiclassical physics a 
singularity arises here. Probably at this stage the black hole has the 
characteristics of an extreme black hole, when the external event horizon 
and internal Cauchy horizon coincide.

As for the processes inside a true singularity in the black hole's 
interior, they can be treated only in the framework of an unified quantum 
theory incorporating gravitation, which is unknown. Thus when we discuss 
any singularity inside a black hole, we should consider the regions with 
the spacetime curvature bigger than $l_{pl}^{-2}$ as physical singularity 
from the point of view of semiclassical physics\footnote{Quantum effects
may manifest themselves in the region with the spacetime curvature smaller 
than $l_{pl}^{-2}$, see [44,45,2].}. About different aspects of quantum 
effects in black holes see also [45,46].

\section{Truly realistic black holes}

So far we discussed the isolated black holes which were formed as the 
result of a realistic gravitational collapse without any assumptions about 
symmetry. Still they are not truly realistic black holes. For the truly 
realistic black holes we should account matter and radiation falling down 
through the event horizon at all times up to infinity (or up to the 
evaporation of a black hole). 

The perturbations at the event horizon which arise as a result of the 
collapse of the non-symmetrical body have a compact support at some 
initial time. Subsequent perturbations, for example the perturbations
which arise from the capture of photons which originate from the relic 
cosmic background radiation, have non-compact support.

We should account also the difference of the curvature of spacetime in the
real Universe from the curvature in the ideal model asymptotically far 
from the black hole (the scattering of the field in these regions is 
responsible for the formation of the late-time power-law radiative tails).
First steps in the investigation of the truly realistic black holes were 
done recently.

Burko [25] studied numerically the origin of the singularity in a simple 
toy model of a spherical charged black hole which was perturbed 
nonlinearly by a self-gravitating spherical scalar field. This field was 
specified in such a way that it had a non-compact support. Namely, it 
grows logarithmically with advanced time along an outgoing characteristic 
hypersurface. It was demonstrated that in this case the weak null Cauchy 
horizon singularity was formed. The null generators of the singularity 
contract with retarded time, and eventually the central spacelike strong 
singularity forms. Thus in this case the casual structure of the 
singularity is the same as in the case of the perturbations with a compact
support at some initial time. Of course, this example is very far to be 
a realistic one.

In another work Burko [26] demonstrated numerically that the scalar field 
can be chosen along an outgoing characteristic hypersurface in such a way
that only spacelike strong singularity forms. The scalar field has a
non-compact support in this case.

It is an open question whether these results hold also for rotating black 
holes, and what would be a result in the case of a realistic source of 
perturbations for realistic black holes.

I want to do the following remark. As I mentioned in Section 5, when 
inside a black hole we come to the singularity close enough, where the 
spacetime curvature reaches $l_{pl}^{-2}$, we should consider this region 
as a singular one from the point of view of semiclassical physics. This 
means that any details of the classical spacetime structure in the 
singular quantum region make no sense. This means that if we are 
interested in the spacetime structure only outside the singular quantum 
region and want to investigate this structure in some definite region at 
the singularity we should take into account radiation coming to the event 
horizon during the restrict period of time $t_0$ only. All radiation which
comes to the border of a black hole later will come to the region under 
consideration inside the quantum singular region and does not influence on
the structure of the spacetime outside it at this place. It is rather easy
to estimate this period $t_0$. It is
\begin{equation}
t_0\approx 3\cdot 10^6\mbox{sec}\left(\frac{M}{10^9M_{\odot}}\right).
\label{q7}
\end{equation}

\section{Can one see what happens inside a Black Hole?}

Is it possible for a distant observer to receive information about the 
interior of a black hole? Strictly speaking, this is forbidden by the very 
definition of a black hole. What we have in mind in asking this question 
is the following. Suppose there exists a stationary or static black hole. 
Can we, by using some device, get information about the region lying 
inside the apparent horizon?

Certainly it is possible if one is allowed to violate the weak energy 
condition. For example, if one sends into a black hole some amount of 
``matter'' of negative mass, the surface of black hole shrinks, and some 
of the rays which previously were trapped inside the black hole would be 
able to leave it. If the decrease of the black hole mass during this 
process is small, then only a very narrow region lying directly inside the
horizon of the former black hole becomes visible.

In order to be able to get information from regions not close the apparent 
horizon but deep inside an original black hole, one needs to change 
drastically the parameters of the black hole or even completely destroy 
it. A formal solution corresponding to such a destruction can be obtained 
if one considers a spherically symmetric collapse of negative mass into a 
black hole. The black hole destruction occurs when the negative mass of 
the collapsing matter becomes equal to the original mass of the black 
hole. In such a case an external observer can see some region close to the
singularity. But even in this case the four-dimensional region of the 
black hole interior which becomes visible has a four-dimensional spacetime
volume of order $M^{4}$. It is much smaller than four-volume of the black 
hole interior, which remains invisible and which is of order $M^{3}T$, 
where $T$ is the time interval between the black hole formation and its 
destruction (we assume $T\gg M$). The price paid for the possibility of 
seeing even this small part of the depths of the black hole is its 
complete destruction.

Does this mean that it is impossible to see what happens inside the 
apparent horizon without a destructive intervention? We show that such a 
possibility exists (Frolov and Novikov [47,48]). In particular, in these
works we discuss
a gedanken experiment which demonstrates that traversable wormholes (if 
only they exist) can be used to get information from the interior of a 
black hole practically without changing its gravitational field. Namely, 
we assume that there exist a traversable wormhole, and its mouths are 
freely falling into a black hole. If one of the mouths crosses the 
gravitational radius earlier than the other, then rays passing through the
first mouth can escape from the region lying inside the gravitational 
radius. Such rays would go through the wormhole and enter the outside 
region through the second mouth, see details in [2,49].

\vspace{0.5cm}

{\bf Acknowledgments}

This paper was supported in part by the Danish natural Science Research 
Council through grant No. 9701841 and also in part by Danmarks 
Grundforskningsfond through its support for establishment of the TAC.

\end{document}